\documentclass[10pt,twocolumn,letterpaper]{article}
\usepackage{iccv}
\usepackage{times}
\usepackage{epsfig}
\usepackage{graphicx}
\usepackage{amsmath}
\usepackage{amssymb}

\usepackage{booktabs}
\usepackage[dvipsnames]{xcolor}
\usepackage{multirow}
\usepackage{hhline}
\usepackage{mathtools}
\usepackage{siunitx}
\usepackage{array}
\usepackage{tabularx}
\usepackage{lipsum}
\usepackage{cite}
\usepackage{url}

\usepackage{makecell}

\newcolumntype{C}[1]{>{\centering\arraybackslash}m{#1}}

\DeclarePairedDelimiter{\norm}{\lVert}{\rVert}

\usepackage[pagebackref=true,breaklinks=true,letterpaper=true,colorlinks,bookmarks=false]{hyperref}

\iccvfinalcopy 

\def\iccvPaperID{7283} 

\ificcvfinal\fi

\begin{document}

\title{Lightweight Hybrid Video Compression Framework Using Reference-Guided Restoration Network}

\author{Hochang~Rhee$^1$, Seyun~Kim$^2$, Nam~Ik~Cho$^1$\\
$^1$Seoul National University, Seoul, Korea\\
$^1$Dept. of Electrical and Computer Engineering, INMC\\
$^2$Gauss Labs Inc.\\
{\tt\small hochangispl@gmail.com, seyun.kim@gausslabs.ai, nicho@snu.ac.kr}}

\maketitle
\ificcvfinal\thispagestyle{empty}\fi

\begin{abstract}
Recent deep-learning-based video compression methods brought coding gains over conventional codecs such as AVC and HEVC. However, learning-based codecs generally require considerable computation time and model complexity. In this paper, we propose a new lightweight hybrid video codec consisting of a conventional video codec (HEVC or VVC), a lossless image codec, and our new restoration network. Precisely, our encoder consists of a conventional video encoder and a lossless image encoder, transmitting a lossy-compressed video bitstream along with a losslessly-compressed reference frame. The decoder is constructed with corresponding video/image decoders and a new restoration network, which enhances the compressed video in two-step processes. In the first step, a network trained with a  video dataset restores the details lost by the conventional encoder. Then, we further boost the video quality with the guidance of a reference image, which is a losslessly compressed video frame. The reference image provides video-specific information, which can be utilized to better restore the details of a compressed video. Experimental results show that when the HEVC is employed as a baseline in our framework, the overall coding gain is comparable to recent top-tier neural codecs while requiring much less encoding time and lower complexity. When combined with the VVC, our method brings significant gain over the VVC, thus achieving state-of-the-art coding performance.
\end{abstract}

\section{Introduction}
\label{sec:introduction}
Despite the development of efficient communication hardware and protocols, it is still hard to meet the bandwidths for transmitting the ever-increasing number of videos. Hence, better and better video compression is required to transmit the increasing number of videos, along with the growing frame sizes and rates. For efficient use of limited bandwidth, finding a balance between video quality and transmission rate is important. In the mathematical term, this is expressed as a rate-distortion trade-off $R+\lambda D$, where $R$ and $D$ denote bitrate and distortion, respectively, and $\lambda$ represents a balancing parameter.


Deep learning has been adopted to enhance the performances of computer vision and image processing tasks, including video compression. Some recent video compression methods~\cite{dvc, fvc, elf-vc, c2f, rafc, dcvc, interframe_compression, scalespace, lu2020, hu2020} maintain the predict-transform architecture of the conventional methods, where they replace the hand-crafted modules with powerful deep neural networks (DNNs). The generalizability and expressive abilities of DNNs are utilized to generate precise motion vectors and compact transforms. These methods are optimized to directly minimize $R+\lambda D$ and achieve superior performance compared to conventional codecs. Although they show impressive performance in terms of the RD curve, they usually require large networks and considerably long computation times for encoding and decoding. For instance, the state-of-the-art method C2F~\cite{c2f} requires about 20 times more decoding time on a GPU than the HECV on a CPU. Also, existing deep-learning methods~\cite{dvc, fvc, dcvc} use complex components such as optical flow and non-local attention, impeding efficient implementation and practical use.

\begin{figure*}[t]
    \centering
    \includegraphics[width=0.9\linewidth]{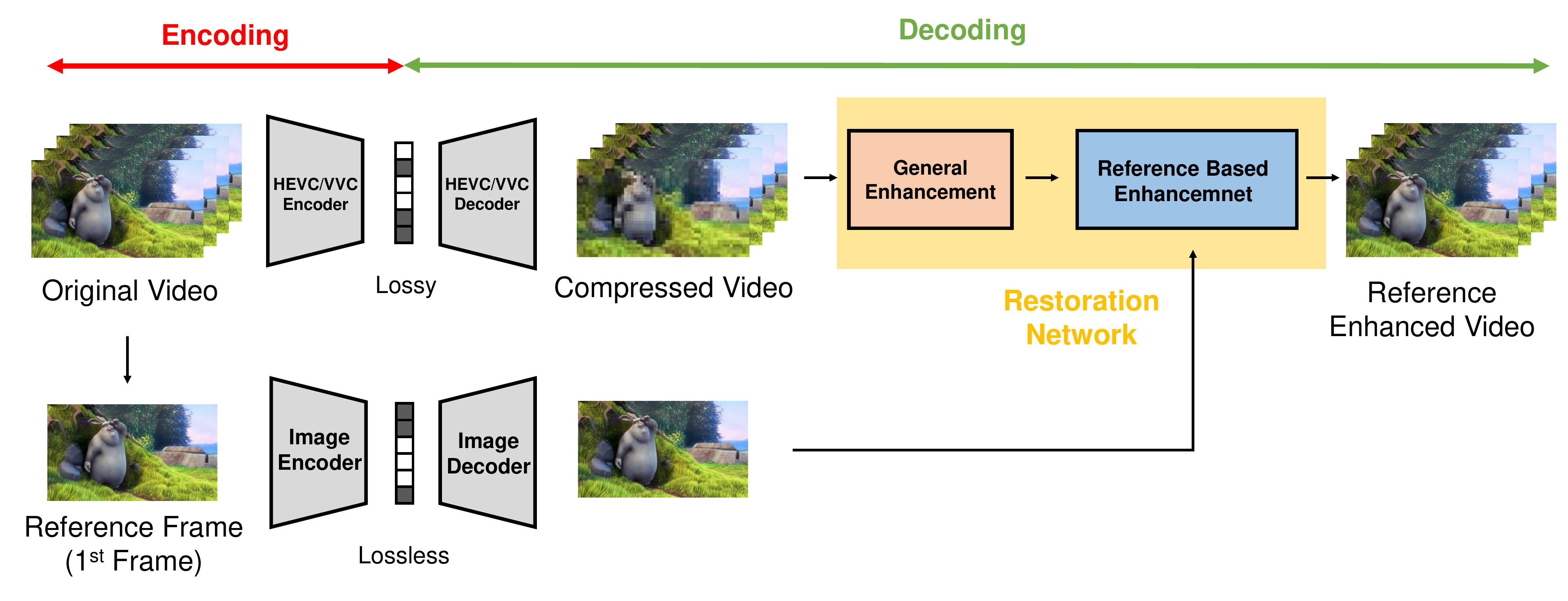}
    \caption{The framework of the proposed method. In the encoding step, the original video is initially compressed using HEVC or VVC. In addition, we losslessly encode a reference frame, which is the first frame or other selected frames of the original video. The decoder receives the conventional compressed video and passes it to the restoration network, which performs a two-step enhancement. The first step restores the video based on the learning from training data, while the second step utilizes the reference frame for video-wise enhancement.}
    \label{fig:framework}    
\end{figure*}

Therefore, in this work, we propose a novel hybrid framework illustrated in Fig.~\ref{fig:framework}, where the encoder is a joint of a conventional video encoder and a lossless image encoder, and the decoder consists of corresponding decoders and a deep restoration network. As shown, the encoder transmits two bitstreams of 1) HEVC or VVC-compressed video and 2) losslessly compressed reference image, which is usually the first frame of a given video shot or selected frame(s). The decoder reconstructs the compressed video and the reference image using the corresponding decoders. Then, the restoration network is applied to enhance the quality of the compressed video. In the enhancement process, the reference image provides video-specific information for better restoring the details.

The restoration network enhances the video in two steps. In the first step, we train a network to learn and restore the details lost by the video encoder. Relying on the generalization power of a DNN trained by a video dataset, we recover the details lost by the conventional codec. This step can be considered a compression artifact reduction network like \cite{nlconvlstm, boosting, artifact_temporal, framewise, highfrequency}. However, the first step has the limitation that the restoration is based on general datasets and does not utilize any video-specific information. Hence, we further enhance the video with the guidance of a reference image. The reference image is an uncompressed frame, which drives the enhancement process to be video-specific, leading to improved restoration for both background and textures. To better utilize the reference image, we let the network derive a confidence map indicating which pixels to use for the enhancement. Through the confidence map, we can efficiently exclude parts of the reference image that are uncorrelated to the current frame, \textit{i.e.}, components that may rather increase the distortion due to the misguided information.

In contrast to prior works that require complex structures, the proposed method requires only simple convolutional neural networks (CNNs) for restoration and a single deformable convolution~\cite{deformable} to align the features of the current and reference frame. Experiments show that the proposed method achieves comparable performance for high-resolution benchmarks while requiring practical decoding times when the HEVC is employed in Fig.~\ref{fig:framework}. For the case of using the VVC, we still obtain significant coding gains over the VVC, leading to higher performance compared to other neural codecs. In summary, the main contributions are as follows:

\begin{itemize}
\item We propose a novel hybrid video compression method, consisting of a conventional video codec, a lossless image codec, and a reference-guided restoration network.

\item The compressed video is enhanced in a two-step procedure, where the first step restores the details lost by conventional video encoding, and the second step uses video-specific information from the reference image for further enhancement.

\item Our method requires only 38\:ms for decoding a 2K frame on NVIDIA 1080 Ti while achieving comparable or better performance than top-tier methods. Hence, we expect our decoder can be performed in real-time on mediocre GPUs. Encoding is also conducted fast on a CPU because we use a hybrid of a conventional video encoder and a lossless image encoder.
\end{itemize}

\begin{figure*}[t]
    \centering
    \includegraphics[width=1.0\linewidth]{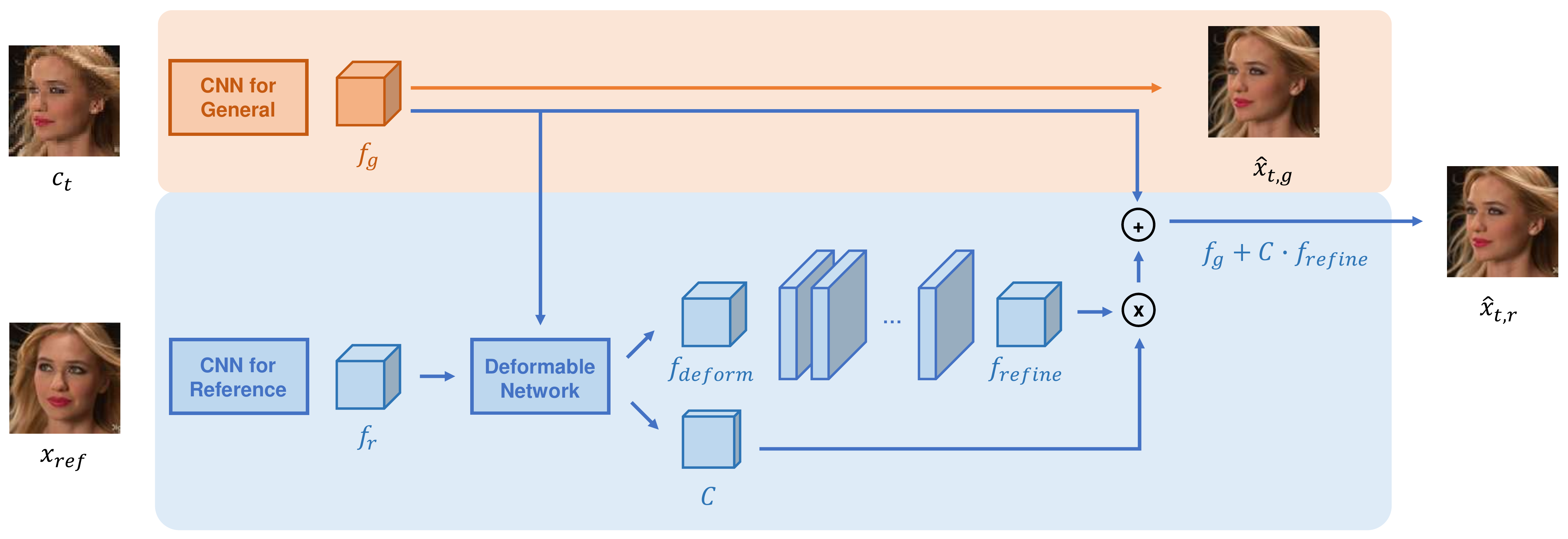}
    \caption{Architecture of the proposed restoration network. The enhancement is progressed in two steps, where the red-colored components indicate the general enhancement (first step) and the blue ones indicate the reference-based enhancement (second step). We receive a compressed frame $c_t$ and use the reference frame $x_{ref}$ to generate an enhanced frame $\hat{x}_{t,r}$.}
    \label{fig:architecture}    
\end{figure*}

\section{Related Works}
\label{sec:related_works}


\subsection{Learning-based Video Compression}
Early learning-based video compression methods \cite{deepcoder, cupartition, learned_entropy} proposed to use DNNs in the places of the conventional codec elements, such as intra-prediction, entropy coding, and mode decision. Some recent methods \cite{dvc, fvc, elf-vc, c2f, rafc, dcvc, interframe_compression, scalespace, lu2020, hu2020, lsvc, feng2020, mlvc, lombardo2019, lu2020content, mentzer2021towards, video_interpolation, hierarchical_quality, multi-modality} focused on reducing the temporal redundancy by improving the elements such as motion estimation, motion compensation, and transform. For example, DVC~\cite{dvc} computes the motion vectors and residuals using DNNs in the image space. It adopts existing optical flow networks to obtain precise optical flows, leading to accurate motion vectors. FVC~\cite{fvc} is an improved work of DVC, where the intermediate components are computed in the feature space rather than the image space. It utilizes stacks of deformable convolution and non-local attention to align the frames. C2F~\cite{c2f} enhances the motion compensation by processing in a coarse to fine manner. It introduces hyperprior-guided mode prediction schemes that can predict the optimal coding modes. Although these works achieve improved performance compared to conventional codecs, they need significantly large encoding/decoding times on GPUs since they require expensive components and large network models.


\subsection{Video Compression Artifact Reduction}
Video compression artifact reduction aims to recover high-quality videos from low-quality compressed videos. They are usually post-processing methods on the decoder side, without modifying the codecs. Existing methods can be categorized into three groups based on the number of input frames and their domain knowledge: image-based, single-frame, and multi-frame approaches. Image-based approaches~\cite{arcnn, dncnn, dpw-sdnet} improve the quality of each frame without utilizing prior knowledge of video compression algorithms. Single-frame approaches~\cite{vrcnn, qe-cnn, partitionaware,artifact_kim} utilize spatial information from a single input frame. Multi-frame methods~\cite{nlconvlstm, boosting, pstqe, mfqe, mfqe2, dfkn, spatiotemporal} utilize temporal information by incorporating adjacent frames. Our method is similar to video compression artifact reduction in that we do not modify the conventional codec and only process on the decoder side. However, while artifact reduction methods enhance quality without additional bitrate, our method utilizes external information at an additional cost for better quality enhancement. Therefore, we compare our method with video compression methods instead of artifact reduction methods.

\section{Method}
\label{sec:method}

\subsection{Overview}
\label{subsec:overview}
Let $\mathbf{x}=\{x_t\}_{t=1,\cdots,T}$ denotes an uncompressed video sequence of length $T$, where $x_t$ indicates the $t$-th frame. We initially use conventional video codec (HEVC or VVC) to compress the video $\mathbf{x}$. Then, we add a losslessly compressed reference frame, denoted as  $x_{ref}$. The first frame is typically used as the reference for simplicity, but it can be replaced by others when there is a significant scene change or when the shot is very long. At the decoder, we first use HEVC or VVC decoder to obtain the compressed video sequence $\mathbf{c}=\{c_t\}_{t=1,\cdots,T}$. Then we feed the $c_t$ and $x_{ref}$ to the restoration network to generate a visually enhanced frame $\hat{x}_t$. The overall architecture of the restoration network is illustrated in Fig.~\ref{fig:architecture}, which performs the two-step enhancement. 


\subsection{General Enhancement}
\label{subsec:general_enhancement}
The first step of our method, which we refer to as \textit{general enhancement}, corresponds to the red-colored components in Fig.~\ref{fig:architecture}. This step enhances the visual quality of the compressed frames by learning the general behavior of HEVC or VVC. We first receive a compressed frame $c_t$ and feed it to a CNN to generate an intermediate feature $f_{g}$. The intermediate feature is fed to the decoding CNN to produce an enhanced frame $\hat{x}_{t,g}$. The networks are trained with the L2 loss between the ground truth (uncompressed) frame and the enhanced frame:

\begin{equation}\label{eq:step1_loss}
L_{general} = \norm{x_t-\hat{x}_{t,g}}_{2}.
\end{equation}

For the training, we compress videos in a dataset using HEVC or VVC and prepare pairs of uncompressed and compressed frames. Through this process, the network learns to restore the details that conventional codecs generally lose. 

Most video encoding and restoration method use a set of consecutive frames as input for exploiting temporal correlation. Hence, it would be natural to use several consecutive compressed frames to obtain an enhanced frame. For example, we may design a framework to obtain an enhanced frame $\hat{x}_{t,g}$ from $c_{t-1}$, $c_{t}$, and $c_{t+1}$ as input frames. However, it is necessary to have precise motion information or accurately align the input frames for a successful result, which requires expensive components such as optical flow or non-local attention. Employing these elements requires too much memory and computation cost, and the errors in motion estimation and alignment would rather deteriorate the performance. Hence, we use only a single corresponding compressed frame $c_{t}$ for obtaining $\hat{x}_{t,g}$, which obviates complex motion compensation systems and error propagation and also adds practicality in computation time.

\subsection{Reference based Enhancement}
\label{subsec:reference enhancement}
In this section, we give details of the \textit{reference-based enhancement} step, which corresponds to the blue-colored components in Fig.~\ref{fig:architecture}. This step aims to enhance the visual quality of a frame based on the provided video-specific information. In general, we utilize the losslessly-compressed first frame of the video as the reference frame $x_{ref}$. For special cases such as dynamic videos with scene changes, we supply additional reference frames, which will be explained later.

Using the reference image can be both beneficial and problematic. The distortion can be enlarged if irrelevant components from the reference image are used. In this sense, finding the appropriate match between the reference and the current image is essential. For example, previous reference-based super-resolution methods adopted optical flow or non-local attention to derive an explicit dense correspondence map between the images \cite{taskdecoupled,c2_matching}. However, deriving the map is too expensive to be practical for video applications.

Since obtaining a dense correspondence map needs excessive computation, we adopt the deformable convolution, which we believe is suitable for our framework for two reasons. First, there is no burden on the memory capacity since correspondence with only a fixed number of components (e.g., 9 components in the case of 3 $\times$ 3 filters) is found. In addition, whereas explicit matching finds components based on similarity, deformable convolution obtains correspondence based on the contribution to the enhancement. In other words, deformable convolution finds the best matches that effectively minimize the loss, \textit{i.e.}, compression distortion in our case.

We perform alignment between $c_t$ and $x_{ref}$ with deformable convolution in the feature space. Initially, $c_t$ and $x_{ref}$ are fed to the corresponding CNNs, which generate $f_{g}$ and $f_{r}$, respectively. Then, the deformable network receives $f_{g}$ and $f_{r}$ and generates two outputs: enhanced feature $f_{deform}$ and a confidence map $C$. The confidence map $C$ is introduced to allow the network to exclude uncorrelated components of the reference. Formally, the process of the deformable network is expressed as:

\begin{equation}\label{eq:deformable_offsetmask}
o, m  = \mathbf{F}(f_{g}, f_{r}),
\end{equation}

\begin{equation}\label{eq:deformable_conv}
f_{deform}, C = D(f_{g}, o, m),
\end{equation}

\begin{figure*}[t]
    \centering
    \includegraphics[width=1.0\linewidth]{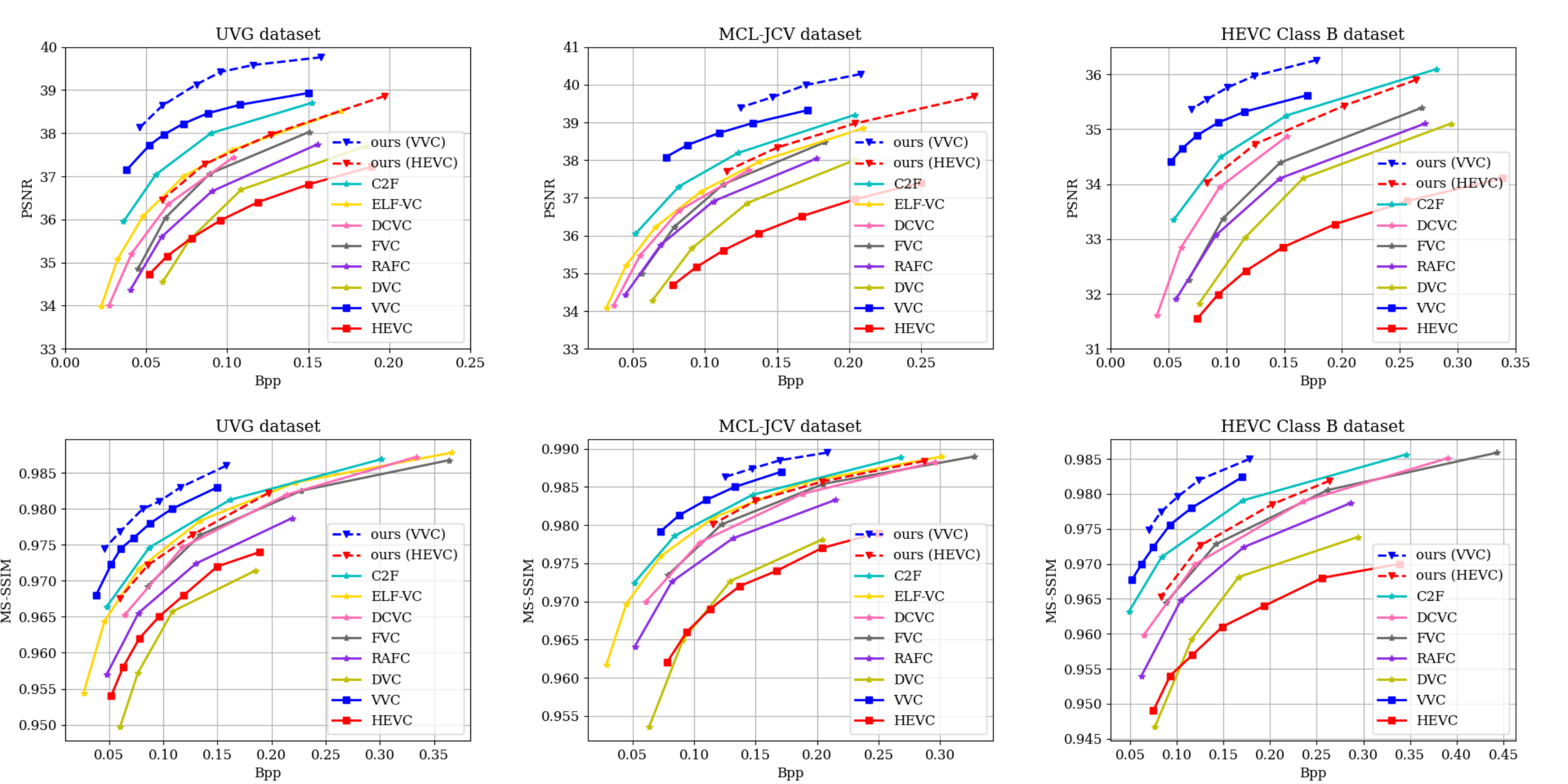}
    \caption{RD curves of PSNR and MS-SSIM on UVG, MCL-JCV, and HEVC Class B datasets.}
    \label{fig:compression_result}
\end{figure*}

\noindent where $o$, $m$, $\mathbf{F}$, and $D$ denote offset, modulation mask, stacks of convolution layers, and deformable convolution, respectively. A large receptive field is necessary to find the correspondence between the current and reference frames. Although stacking deformable convolution can be a straightforward solution, this leads to long and unstable training. Hence, instead, we stack the convolution layers to derive the offset and modulation mask to enlarge the receptive field.

After aligning through the deformable network, we refine $f_{deform}$ through a stack of convolution layers to obtain $f_{refine}$. Then, we use the confidence map derived from Eq.~\ref{eq:deformable_conv} to suppress the weight of the misaligned regions and obtain the final feature as:

\begin{equation}\label{eq:feature}
f_{out} = f_{g} + C \cdot f_{refine}.
\end{equation}

\noindent Since $f_{g}$ contains information for restoring details that HEVC or VVC generally removes, $f_{refine}$ only needs to concentrate on utilizing the video-specific information provided from the reference frame. The final feature is fed to the decoding CNN to generate the final prediction $\hat{x}_{t,r}$. The networks are trained with the L2 loss between the ground truth and the enhanced frame:

\begin{equation}\label{eq:step2_loss}
L_{ref} = \norm{x_t-\hat{x}_{t,r}}_{2}.
\end{equation}

We freeze the networks of the first step when training the reference-based enhancement. This further constrains the networks of the second step to focus on using video-specific information and not consider the behavior of HEVC or VVC. It will be shown that proceeding in this two-step procedure shows superior performance compared to training all the networks together.

\subsection{Multiple Reference Frame}
\label{subsec:multiple reference frame}
In our framework, we typically use the uncompressed first frame as the reference. This method shows great strength for static videos, as the reference frame contributes significantly to the enhancement throughout the entire video. However, the impact of the reference frame is less significant in the case of dynamic videos, particularly in the later parts of the video. In this case, using an additional reference frame at the cost of extra bits is worthwhile from the perspective of RD curve. However, to determine when to supply a new reference frame, we need to answer two questions: 1) which videos benefit from additional reference frames, and 2) at which frame should the new reference frame be supplied? We resolve the aforementioned questions by applying a video scene cut detection tool, PySceneDetect~\cite{pyscenedetect}. We use PySceneDetect to identify dynamic videos with a scene change and supply a new reference frame at the point of the scene change. In this case, reference frames can contribute throughout the whole video.

\section{Experiments}

\subsection{Experimental Setup}

\noindent \textbf{Evaluation Dataset} We validate our method on three 2K resolution benchmark datasets, UVG~\cite{uvg}, MCL-JCV~\cite{mcl-jcv}, and HEVC Class B~\cite{hevc}. The UVG (Ultra Video Group) dataset consists of 7 videos, each containing 600 frames except for the one having 300 frames. The MCL-JCV consists of 30 videos, where each consists of approximately 150 frames. Note that a large portion of videos in UVG is static, whereas MCL-JCV contains a portion of dynamic scenes. HEVC Class B contains 5 videos consisting of 240 to 600 frames. 

\vspace{0.2cm}
\noindent \textbf{Evaluation Metric} We adopt PSNR and MS-SSIM~\cite{ssim} for evaluating video quality, which are commonly used metrics in image/video compression. For evaluating the rate, we use \textit{bits per pixel} (bpp).

\vspace{0.2cm}
\noindent \textbf{Training} We train our network with Vimeo-90K dataset~\cite{vimeo}, which provides 89,800 video clips, each containing seven frames. In addition, we use MCL-JCV and HEVC Class B as additional training sets when evaluating for UVG. In the same context, we use UVG and HEVC Class B as extra data for MCL-JCV and so on. We randomly extract a patch of size 256$\times$256 and apply data augmentation during training. We apply rotation and flipping as augmentation. AdamW optimizer~\cite{adamw} is used for the training with a learning rate of 1$\times10^{-4}$. We set the batch size as four and train over five epochs, each for the first and second steps. The training takes 48 hours and 72 hours respectively for the first and second steps when trained on a GeForce RTX 2080 Ti.

\vspace{0.2cm}
\noindent \textbf{Implementation Detail} Since our method is a restoration task, maintaining the image resolution is favorable. Hence, we use only fully convolutional layers and exclude components that downscale the resolution, such as pooling and strided convolution. For the HEVC, we use \textit{ffmpeg} under the \textit{medium} setting. For the VVC~\cite{vvc}, we use VVenC~\cite{vvenc} and VVdeC~\cite{vvdec} which are fast and efficient H.266/VVC software implementations. We use VVC under the \textit{medium} setting. For the compression of the reference image, we adopt a fast and efficient image codec JPEG-XL~\cite{JPEG-XL}. Further details of the network architecture, \textit{ffmpeg} command line, VVenC/VVdeC command line, and PySceneDetect~\cite{pyscenedetect} parameters are provided in the Supplementary Material.

\subsection{Compression Result}
In this section, we compare our method with the state-of-the-art methods including, C2F~\cite{c2f}, ELF-VC~\cite{elf-vc}, DCVC~\cite{dcvc}, RAFC~\cite{rafc}, FVC~\cite{fvc}, and DVC~\cite{dvc}. Conventional codecs include HEVC~\cite{hevc} and VVC~\cite{vvc}, where the preset is set as \textit{medium}. 
Regarding side information for sending a losslessly compressed reference, it needs an average of 5.30 bpp for an image, which is 0.0088 bpp for a 600-frame video.

\begin{table}[]
\centering
\caption{BDBR(\%) results when compared with HEVC \textit{medium} setting. Negative values indicate the bit-rate saving compared to HEVC.}
\resizebox{1.0\columnwidth}{!}{%
\begin{tabular}{c|ccccc}
\hline
Method       & FVC   & DCVC  & ELF-VC & C2F   & Ours (HEVC\:/\:VVC)  \\ \hline
UVG          & $-36.72$ & $-42.72$ & $-51.54$  & $-64.78$ & $-54.27$ / $-90.06$ \\ 
MCL-JCV      & $-46.22$ & $-51.03$ & $-54.61$  & $-64.29$ & $-58.54$ / $-98.98$  \\ 
HEVC Class B & $-50.67$ & $-59.78$ & -       & $-75.38$ & $-74.27$ / $-95.60$ \\ \hline
\end{tabular}%
}
\label{table:compression_result}
\end{table}

Fig.~\ref{fig:compression_result} presents the RD curves for the benchmark datasets. When applied to VVC, our method shows the best performance on all datasets in terms of PSNR/MS-SSIM. When applied to HEVC, our method improves the coding performance to be comparable to the second-best neural codec ELF-VC. For the RD curves regarding MS-SSIM, we fine-tune our networks with the MS-SSIM loss. The MS-SSIM results show a similar trend to that of PSNR, demonstrating that our method performs the best when applied to VVC and is comparable to other neural codecs when applied to HEVC. Table~\ref{table:compression_result} provides the results of BDBR~\cite{bdbr} compared to HEVC. We can observe that our method is comparable to others when applied to the HEVC and also brings large performance gains when applied to the VVC. In the case of HEVC, although our method is the second-best, it offers advantages regarding running time and model size, which will be explained in the next section.

\subsection{Running Time and Model Complexity} 
\label{subsec:runtime}
In Table~\ref{table:time_result}, we report the encoding/decoding times, model parameters, and GPU settings for various video compression methods. The times are measured for a 2K video with the bpp around 0.2. We measure the time of our method on NVIDIA 1080 Ti to demonstrate that our fast speed does not come from strong GPU powers. As can be seen, our method requires only 38\:/\:54\:ms for decoding, which includes the time for the HEVC\:/\:VVC decoder (13\:/\:29\:ms), reference frame decoding (2\:ms), and the restoration network (23\:ms). Decoding the reference frame using JPEG-XL actually requires 0.64\:s, but since its decoding needs to be performed only once for a given video shot, the running time becomes negligible as the shot length gets longer. In this case, we assume a 300-frame video and measure the reference frame decoding time. 

Our method is the fastest among the learning-based methods while requiring the least model parameter. Regarding the decoding time, our method is 7.7\:/\:5.4 times faster than the state-of-the-art method C2F and achieves 26\:/\:18 FPS, which is near real-time. Nevertheless, our method only requires a model size of 3.7\:M, which is 3 times smaller than the ELF-VC (11\:M). Whereas previous methods introduce heavy models to reduce the temporal redundancy, we leave this complex procedure to conventional codecs. Our method can solely focus on restoration, which can be designed without heavy and complicated components.

\begin{table}[]
\centering
\caption{Comparison of encoding/decoding time, model parameter, and GPU for video codecs. * indicates that the time was measured in CPU. The model parameter is left blank if it was not stated in the corresponding paper.}
\resizebox{1.0\columnwidth}{!}{%
\begin{tabular}{c|c|c|c|c}
\hline
Method & \thead{Encoding \\ Time (ms)} & \thead{Decoding \\ Time (ms)} & Parameter (M) & GPU            \\ \hline
HEVC~\cite{hevc}     & 71*   & 13*        & -         & -              \\ 
VVC~\cite{vvc}     & 2629*   & 29*        & -         & -              \\ 
DVC~\cite{dvc}       & 667  & 460       & 11        & 1080 Ti              \\
FVC~\cite{fvc}       & -    & 548       & 26        & 2080 Ti \\ 
DCVC~\cite{dcvc}     & -    & 857       & -         & P40            \\ 
ELF-VC~\cite{elf-vc} & 200  & 83        & 11        & Titan V \\ 
C2F~\cite{c2f}       & -    & 293       & -         & 2080 Ti \\ 
Ours (HEVC)          & 86*   & 38        & 3.7       & 1080 Ti \\
Ours (VVC)           & 2644* & 54        & 3.7       & 1080 Ti \\ \hline
\end{tabular}%
}
\label{table:time_result}
\end{table}

\begin{figure}
    \centering
    \begin{tabular}{c}
        \includegraphics[width=0.98\columnwidth]{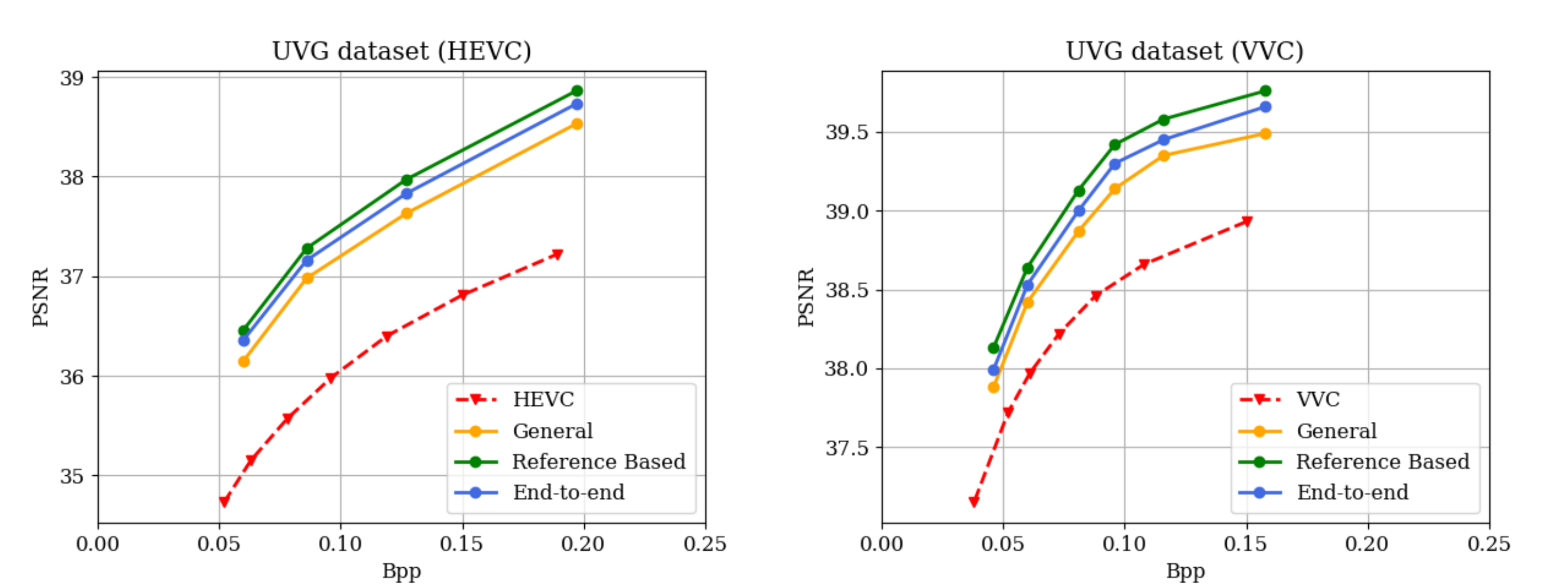}
    \end{tabular}
	\caption{Result of each step in our method applied to HEVC and VVC. The results are evaluated for the UVG dataset. {\em End-to-end} refers to a one-step enhancement method, where all the networks are trained simultaneously.}
	\label{fig:step_analysis}
\end{figure}

\begin{figure*}[t]
    \centering
    \begin{tabular}{c}
        \includegraphics[width=0.9\linewidth]{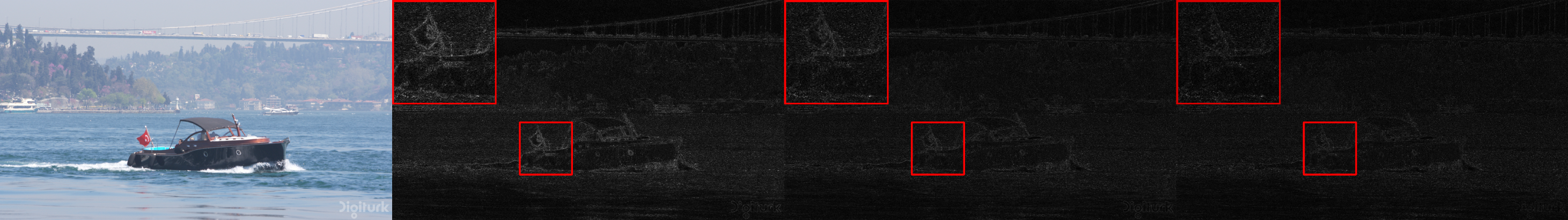} \\
        \includegraphics[width=0.9\linewidth]{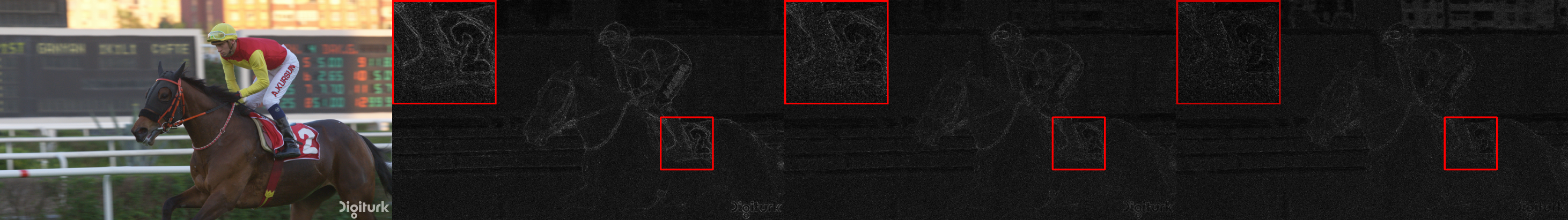} \\
        Raw  \hspace{3cm}  HEVC \hspace{3cm}  Step 1 \hspace{3cm}  Step 2 \\
    \end{tabular}
	\caption{Each row shows the results in the following order: raw, residuals of HEVC compressed, general-enhanced, and reference-based enhanced frames. In the case of compressed frames, the residual from the raw frame is presented for better understanding. The difference is scaled for better visualization. Best viewed in PDF version.}
	\label{fig:step_visualization}    
\end{figure*}

\begin{figure}
    \centering
    \begin{tabular}{c}
        \includegraphics[width=0.95\linewidth]{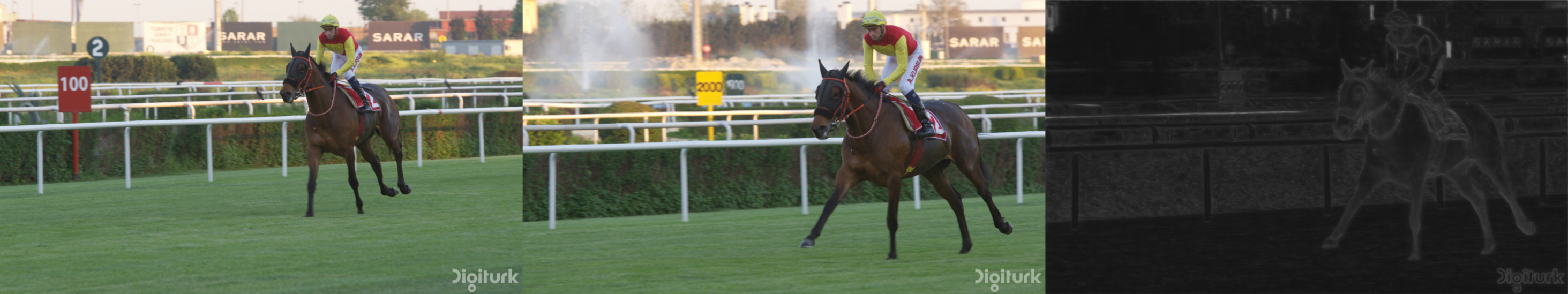} \\
        \includegraphics[width=0.95\linewidth]{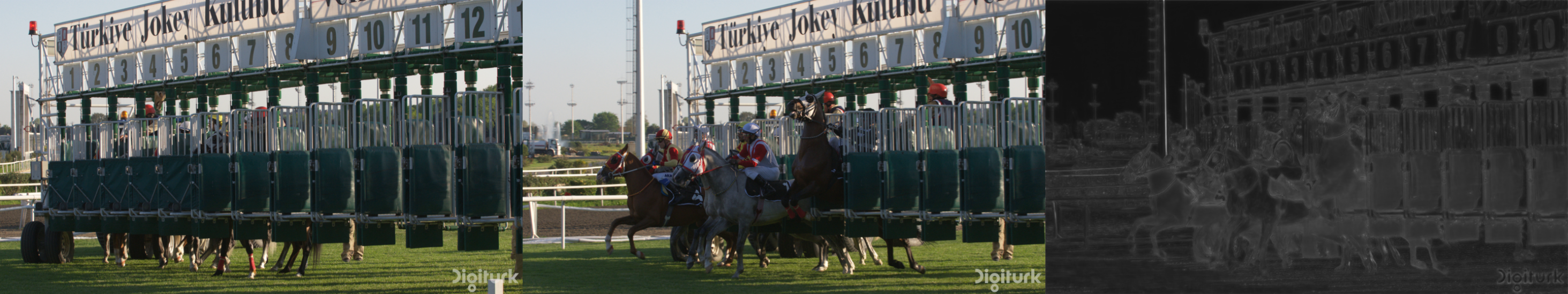} \\
        \includegraphics[width=0.95\linewidth]{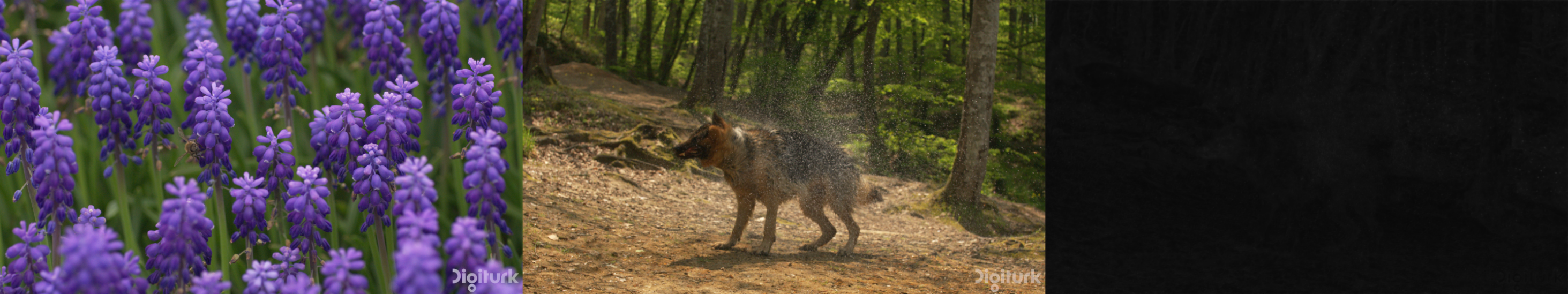} \\
        \:Reference\: \hspace{1.0cm} \:\:\:\:Input\:\:\:\: \hspace{1.0cm} Confidence 
    \end{tabular}
	\caption{Visualization of the reference frame, one of the input frames, and confidence map. The first and second rows show the cases when a relevant reference frame is provided. The third row shows when an irrelevant reference is given. Best viewed in PDF version.}
	\label{fig:confidencemap}    
\end{figure}

\subsection{Step Analysis}
In this section, we analyze the contribution of each step in our method as summarized in Fig.~\ref{fig:step_analysis}.  Our analysis shows that the general enhancement of our method provides an average gain of 1.27 dB and 0.50 dB compared to HEVC and VVC for UVG. Compared to HEVC, the impact of general enhancement is reduced in the case of VVC, likely due to VVC's already strong baseline performance. Compared to general enhancement, we observe that reference-based enhancement shows an average gain of 0.32 dB and 0.25 dB. It can be inferred that the contribution of the reference frame to the enhancement is similar regardless of the strength of the base conventional codec. In conclusion, we can observe that each step has a valid contribution, where the first step restores most of the details and the second step uses video-specific information for further improvement.

In addition, we demonstrate that our two-step enhancement strategy is effective compared to the end-to-end trained method. The end-to-end training means that no network is frozen, and all the networks are trained with the loss of $L = L_{general} + L_{ref}$. Considering the result of the end-to-end training in Fig.~\ref{fig:step_analysis}, we observe a 0.12 dB and 0.13 dB performance drop compared to our full model. We interpret that freezing networks allow each step to focus on their function, \textit{i.e.}, the general enhancement enhances a frame based on the property of HEVC\:/\:VVC, and the reference-based enhancement does on the video-specific information.

Fig.~\ref{fig:step_visualization} presents the visualization result of each step. It can be seen that HEVC usually removes information near edges and textures. Learning the behavior of HEVC from a video dataset, the first step of our method significantly reduces the overall distortion, especially in the texture areas. Finally, the distortion is further decreased when the network is supplied with a reference frame in the second step.

\begin{figure}
    \centering
    \begin{tabular}{c}
        \includegraphics[width=1.0\columnwidth]{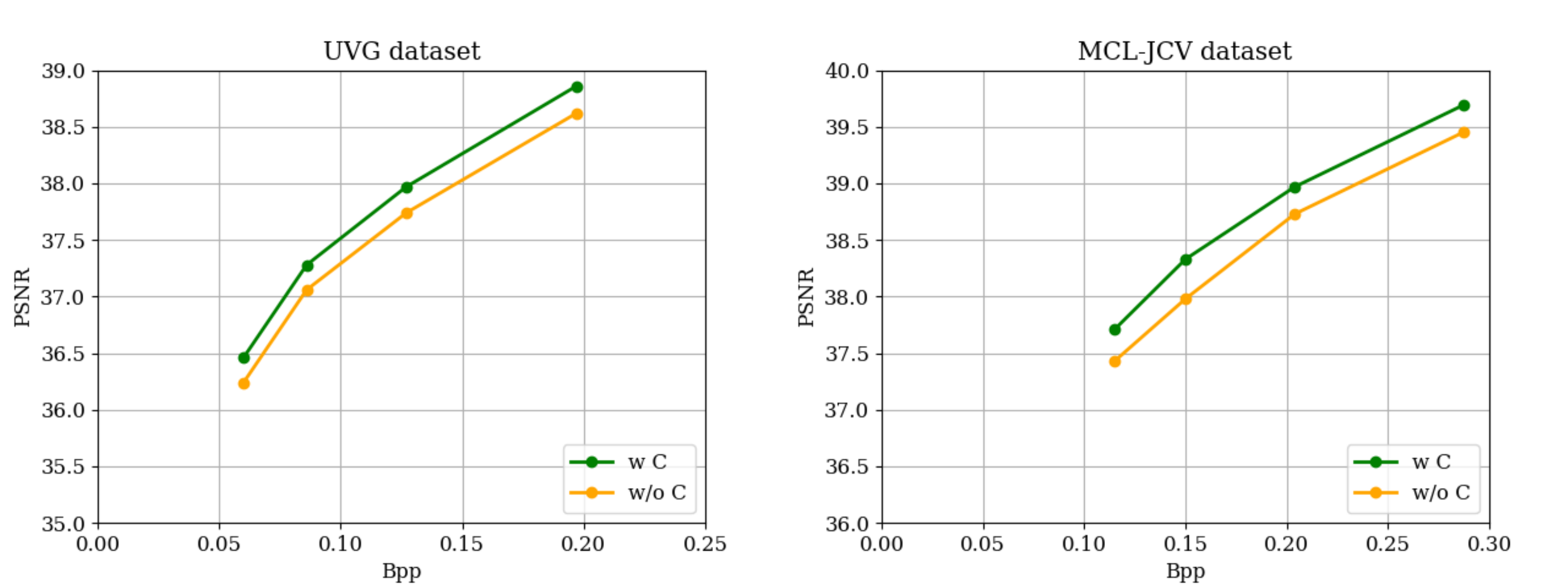}
    \end{tabular}
	\caption{Ablation study on the confidence map. Results are reported for UVG and MCL-JCV datasets when our method is applied to HEVC.}
	\label{fig:conf_ablation}    
\end{figure}

\subsection{Confidence Map}
Fig.~\ref{fig:confidencemap} visualizes the confidence map given the reference and input frame. The first and second rows exhibit the cases when the reference and input frame have relevant features. It should be noted that the confidence map is not a similarity map, \textit{i.e.}, the purpose of the confidence map is not to find and indicate relevant contents. Instead, its role is to specify components that can benefit from video-specific information. We can observe that the confidence map generally has high values near edges and low values in smooth regions. Hence, the video-specific information highly contributes to enhancing edges instead of smooth regions. This is reasonable since the information of smooth regions is not much lost during compression; therefore, not much is left for the reference frame to restore. The third row of Fig.~\ref{fig:confidencemap} shows the case when the reference and input frame have irrelevant features. In this case, the confidence map suppresses almost all the components since they may provide misguided information. Fig.~\ref{fig:conf_ablation} demonstrates the contribution of the confidence map. It can be easily noticed that employing the confidence map boosts the performance, where an average of 0.23 dB and 0.22 dB is increased.

\subsection{Alignment Comparison}
In this section, we demonstrate the contribution of deformable convolution by replacing it with non-local attention and optical flow. For the non-local attention, we use $C^2$-matching~\cite{c2_matching}, which is designed to alleviate the transformation and resolution gap during the alignment process. For the optical flow, we use the widely used PWC-Net~\cite{pwc-net}, which is fast and efficient. Pretrained models are adopted and fine-tuned for our task. Since non-local attention and optical flow only perform matching and no enhancement is done, we stack additional convolution layers after the matching to perform the enhancement. 

The above-mentioned methods are compared in Table~\ref{table:alignment}. PSNR is evaluated for the UVG dataset when the rate is 0.2 bpp. As can be seen, deformable convolution shows superiority in many aspects, specifically the highest PSNR, most minor parameter usage, and the least time consumption. Optical flow shows the lowest performance since matching is operated in the image space, not the feature space. Non-local attention exhibits a slight performance drop despite the computation cost it requires. We conjecture that the performance drop is introduced since the matching is based on similarity. Similar components do not necessarily lead to better enhancement, \textit{i.e.}, it can be rather sub-optimal in terms of distortion minimization.

\begin{table}[]
\centering
\caption{Comparison of the alignment methods. PSNR is evaluated for the UVG dataset on 0.2 bpp, and the decoding time is measured for a 2K video.}
\resizebox{0.8\columnwidth}{!}{%
\begin{tabular}{l|ccc}
\hline
\hfil Method                           &  PSNR (dB) &  Parameters (M) &  \thead{Decoding \\ Time (ms)} \\ \hline 
Optical Flow~\cite{pwc-net}            &  38.59 &  10.6        &  55        \\
Non-local Attention~\cite{c2_matching} &  38.78 &  10.7        &  98        \\ 
Deformable Convolution                 &  38.86 &  \:\:3.7       &  38        \\\hline
\end{tabular}%
}
\label{table:alignment}
\end{table}

\subsection{Reference Frame Analysis}
\begin{figure}[]
    \centering
    \begin{tabular}{cc}
        \includegraphics[width=0.5\columnwidth]{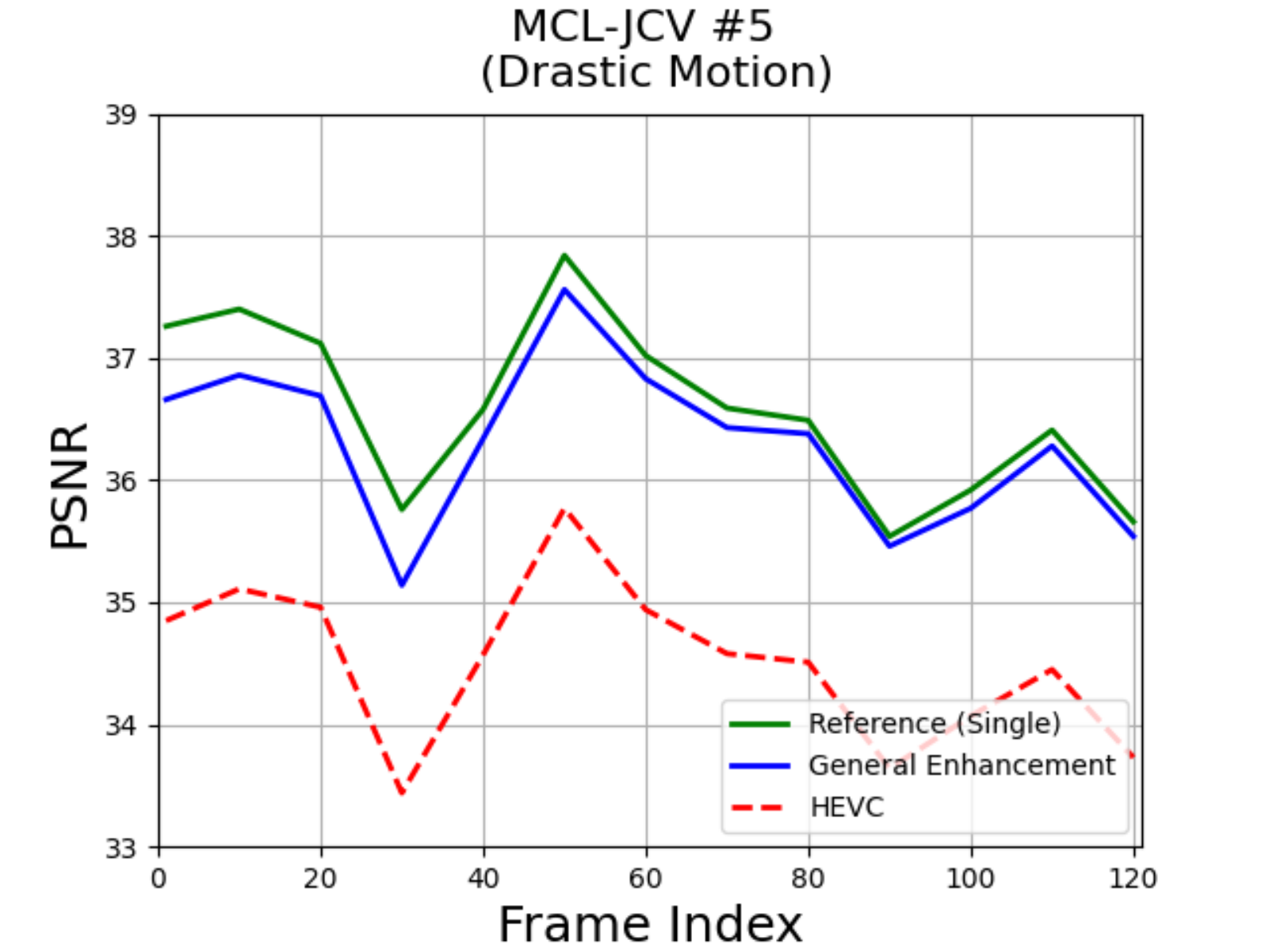} \hspace{-0.5cm} & \includegraphics[width=0.5\columnwidth]{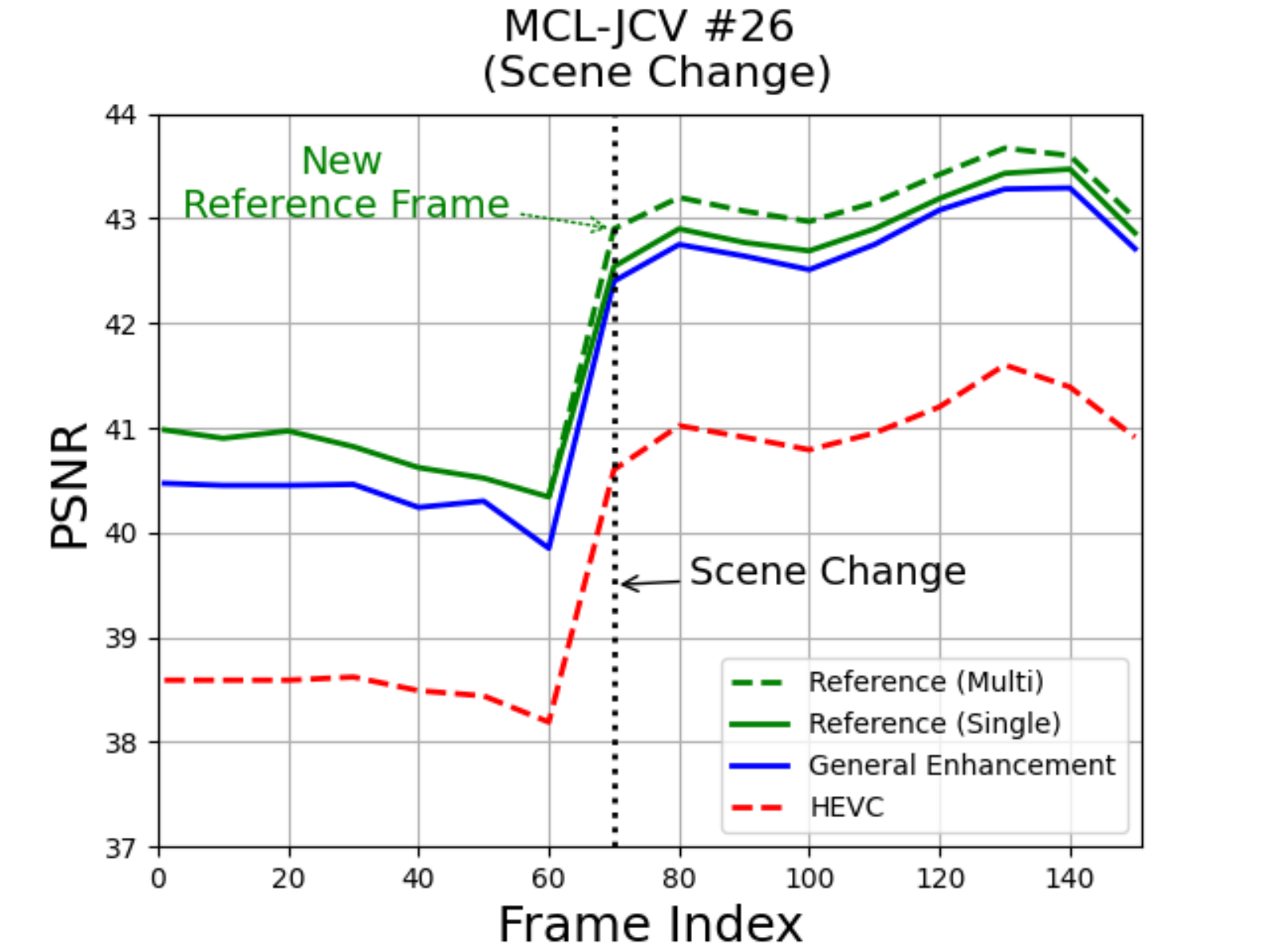}\\
        (a) \hspace{-0.5cm} & (b) \\
    \end{tabular}
	\caption{(a) and (b) illustrates the performance for MCL-JCV \#5 (video with drastic motion) and MCL-JCV \#26 (video with scene change at the 70th frame) respectively.}
	\label{fig:ref_frame_analysis}    
\end{figure}

In Fig.~\ref{fig:ref_frame_analysis}, we present the results for two dynamic videos, one with drastic motion and the other with a scene change. We observe that when there is a high correlation between the reference and the current frame, the reference contributes up to 0.4 dB gain. As the correlation decreases, such as in the later parts of the video or after a scene change, the contribution of the reference frame reduces to 0.1 dB. Although minor, the reference frame still contributes to the enhancement of distant frames. We conjecture that although the reference and distant frames have small correlations in contents, the reference still provides information such as image characteristics. Fig.~\ref{fig:ref_frame_analysis}(b) shows the result for dynamic video with a scene change, where we use additional reference frames in our method. We observe that frames after the scene change significantly benefit from the new reference frame.

In addition, we report the performance when an irrelevant frame is used as the reference frame. Specifically, frames from different videos are used as references. The result is reported in Table~\ref{table:ref_frame_analysis}, where we compare 1) no reference frame (only general enhancement), 2) irrelevant frame, and 3) relevant frame. It can be noticed that using an irrelevant frame is almost equivalent to using no reference frame. Note that the worst possible case of using an irrelevant frame is that the network uses misguided information and rather degrades the visual quality. However, the performance does not drop below \textit{None}, implying that misguided information is not used. This is due to the confidence map, which suppresses the contribution of irrelevant features.

\begin{table}[]
\centering
\caption{PSNR results when a different reference frame is given. \textit{None} is equivalent to using only general enhancement. \textit{Irrelevant} denotes that a frame from other videos is used as the reference. \textit{Relevant} uses the first frame of the video as the reference.}
\vspace{0.2cm}
\resizebox{0.6\columnwidth}{!}{%
\begin{tabular}{c|cc}
\hline
Reference Frame  & UVG      & MCL-JCV  \\ \hline
None   &  38.53 &  39.35 \\ 
Irrelevant    &  38.58 & 39.44 \\ 
Relevant      &  38.86 &  39.69 \\ \hline
\end{tabular}%
}
\label{table:ref_frame_analysis}
\end{table}

\section{Conclusion}
We have proposed a novel hybrid video compression framework, which consists of a conventional video codec, a lossless image codec, and a restoration network. The encoding procedure includes the lossy compression of the video and lossless compression of the reference image. During decoding, the restoration network enhances the visual quality of the compressed video in a two-step process. The first step restores the details lost by the conventional codec in general, and the second step utilizes the reference frame for video-specific enhancement. Comprehensive experiments demonstrate that our method achieves comparable or better performances than top-tier learning-based methods while requiring the shortest decoding time and model complexity.

{\small
\bibliographystyle{ieee_fullname}
\bibliography{egbib}
}

\end{document}